\documentstyle[a4fedele,10pt]{article}

\begin{document}

\newcommand{\vm}{\vspace{0.2cm}}
\newcommand{\vl}{\vspace{0.4cm}}
\newcommand{\per}{\hspace{.2cm}}

\
\vspace{0.5cm}
\begin{flushright}
{\bf HIP-1999-61/TH}
\end{flushright}

\Large
\begin{center}
The Goto-Imamura-Schwinger Term and Renormalization Group

\normalsize \vl

Masud Chaichian\\

\vm

High Energy Physics Division, Department of Physics, \\ 
 University of
Helsinki\\

and \\

Helsinki Institute of Physics\\ P.O. Box 9, FIN-00014
Helsinki,  Finland \\

\vm

and

\vm

Kazuhiko Nishijima \\ 

\vm 

Nishina Memorial Foundation\\ 2-28-45 Honkomagome, Bunkyo-ku, Tokyo
113-8941, Japan

\end{center}

\vspace{2.0cm}

\normalsize

\begin{center} Abstract \end{center}

\vm

In connection with the question of color confinement the origin of the
Goto-Imamura-Schwinger term has been studied with the help of
renormalization group. An emphasis has been laid on the difference
between theories with and without a cut-off.

\section{Introduction}

Field theory is full of ghosts and bugs, and we have to bring
divergences, anomalies and ambiguities under control. Among others we
shall concentrate on the origin of the so-called
Goto-Imamura-Schwinger (GIS) term$^{1,2)}$ in field theory, since it
bears a close connection with the question of color
confinement$^{3-5)}$.

In evaluating the equal-time commutator (ETC) between two local
operators we sometimes encounter a result in conflict with that
obtained by a naive application of the canonical commutation relations
(CCRs). The deviation from the naive expectation is referred to as
the Goto-Imamura-Schwinger (GIS) term hereafter.  Such a term does not
arise, however, when we evaluate the ETC between two fundamental
fields, and it indicates that the origin of the GIS term must be
sought in the definition of the singular product of field operators at
the same space-time point.

In many examples it is possible to find a renormalization group (RG)
equation controlling the GIS term in question, but then the next
question is raised of  how to formulate the initial or boundary
condition for this equation. In the RG approach we introduce running
parameters such as the running coupling constant and they tend to the
bare or nonrenormalized ones in the high energy limit provided that
we introduce a cut-off in the unrenormalized version of the theory as
we shall see in Sec. 2. Then we can introduce boundary conditions in
the high energy limit into the cut-off theory by assuming the CCRs. In
some cases it is possible to formulate the boundary condition
kinematically, namely, without reference to the dynamics of the
system but often it is necessary to refer to the dynamics of the
system by evaluating higher order corrections. In Sec. 3 we shall
illustrate these statements in quantum electrodynamics (QED).  Then,
we find that the origin of the GIS terms may be attributed to one of
the following causes: (1) operator-mixing under
renormalization$^{3,5)}$, (2) non-local character of the product of
field operators at the same space-time point and (3) divergences
induced by lifting the cut-off.  In Sec. 4 we shall proceed to quantum
chromodynamics (QCD) in connection with the question of color
confinement.

\section{Renormalization Group}

In introducing the RG approach$^{6-8)}$ we shall employ the neutral
scalar theory for illustration. We assume the quartic interaction of
the scalar field $\phi(x)$ with the coupling constant $g$. The
unrenormalized Green function is given by

\begin{eqnarray}
G_0^{(n)}(x_1,...,x_n)=\langle 0\vert T\left[\phi^{(0)}(x_1)\cdots
\phi^{(0)}(x_n) \right]  \vert 0\rangle \per ,
\label{2.1}\end{eqnarray}

\noindent where the subscript $0$ and the superscript $(0)$ denote
unrenormalized quantities. The Fourier transform of the renormalized
n-point Green function is denoted by

\begin{eqnarray}
  G^{(n)}(p_1,...,p_n;g(\mu),\mu)\per ,
\label{2.2}\end{eqnarray}

\noindent where $\mu$ denotes the renormalization point defined below
and  $g(\mu)$ the running coupling constant defined at the
renormalization point as seen from
\begin{eqnarray}
(p^2+ m^2)G^{(2)}(p^2;g(\mu),\mu)=1\per , \hspace{4mm}\mbox{for} \per p^2
=\mu^2\per ,
\label{2.3}
\end{eqnarray}
\begin{eqnarray}
&& G_{conn}^{(4)}(p_1,...,p_4;g(\mu),\mu)\nonumber\\
&&= g(\mu)\prod_{i=1}^4G^{(2)}(p^2_i;g(\mu),\mu)\cdot
\Gamma (p_1,\cdots,p_4;g(\mu),\mu)\per ,
\label{2.4}
\end{eqnarray}
\begin{eqnarray}
\Gamma (p_1,\cdots,p_4;g(\mu),\mu)= 1\per , \hspace{4mm} \mbox{for}\per p_i\cdot p_j
= \frac{\mu^2}{3}(4\delta_{ij}-1)\per ,
\label{2.5} 
\end{eqnarray}
\noindent where $G^{(4)}_{conn}$ denotes the 4-point Green function
for connected Feynmann diagrams alone. These are the normalization
conditions for the Green functions and specify the renormalization
point in the Pauli metric. 

The generator of the RG is given by 

\begin{eqnarray}
{\cal D}= \mu\frac{\partial}{\partial \mu}
 +\beta(g)\frac{\partial}{\partial g}\per ,
\label{2.6}
\end{eqnarray}

\noindent and the RG equation for the n-point Green function is given
by 

\begin{eqnarray}
\left[{\cal D} + n\gamma_{\phi}(g)\right] G^{(n)}(p_1,...,p_n;g,\mu)=0\per
,
\label{2.7}\end{eqnarray}

\noindent where we write $g$ for $g(\mu)$ and $\gamma_{\phi}$ denotes
the anomalous dimension of the scalar field $\phi$. For the two-point
Green function or the propagator we may assume the Lehmann
representation$^{9)}$,

\begin{eqnarray}
G^{(2)}(p^2;g,\mu)= \int d\kappa^2 \frac{{\cal \rho}
 (\kappa^2;g,\mu)}{p^2+\kappa^2 -i\epsilon}\per , 
\label{2.8}\end{eqnarray}

\noindent and we have

\begin{eqnarray}
\left[{\cal D}+ 2\gamma_{\phi}(g)\right] {\cal \rho}(\kappa^2;g,\mu)=0\per .
\label{2.9}\end{eqnarray}

\noindent Then Eq. (2.3) in the limit $\mu\rightarrow \infty$
yields

\begin{eqnarray}
\lim_{\mu\rightarrow\infty}(\mu^2+ m^2) G^{(2)}(\mu^2;g,\mu)\nonumber\\ 
\\
= \int d\kappa^2 {\cal \rho}(\kappa^2;g(\infty),\infty) =1 \per ,\nonumber 
\label{2.10}\end{eqnarray}

\noindent in the cut-off theory where $m$ denotes the mass of the
quantum  of the scalar field.

Lehmann's theorem$^{9)}$ on the ETC for the field operator normalized
at $\mu$ readily yields the relation

\begin{eqnarray}
\delta (x_0-y_0)\left[\phi(x;g,\mu), \dot{\phi} (y;g,\mu)  \right]
= i\delta^4 (x-y)\int  d\kappa^2 {\Large \rho} (\kappa^2;g,\mu)\per , 
\label{2.11}\end{eqnarray}

\noindent and Eq. (2.10) then implies that the field operators are
identified with the unrenormalized ones in the limit $\mu\rightarrow
\infty$ since they satisfy the CCR.  At the same time we can show
that $g(\mu)$ also tends to the bare coupling constant $g_0$ in the
same limit.

In order to define the running parameters we introduce

\begin{eqnarray}
R(\rho) = exp(\rho \cal{D})\per ,
\label{2.12}\end{eqnarray}

\noindent where $\rho$ denotes the parameter of the RG, then $R(\rho)$ obeys
the composition law

\begin{eqnarray}
R(\rho_1){\cdot}R(\rho_2)= R(\rho_1+\rho_2)\per ,
\label{2.13}\end{eqnarray}

\noindent and the RG is literally a group identified with $GL(1,R)$.

The running parameters in the scalar theory are defined by

\begin{eqnarray}
\overline{g}(\rho)&=& R(\rho)\cdot g\per ,\\ \overline{\mu}(\rho)&=&
R(\rho)\cdot \mu = \mu~ exp(\rho)\per ,
\label{2.15}\end{eqnarray}

\noindent then we readily obtain

\begin{eqnarray}
R(\rho)~G^{(n)}(p_1,...,p_n;g,\mu)= G^{(n)}(p_1,...,p_n;\overline{g}(\rho),
\overline{\mu}(\rho))\per .
\label{2.16}\end{eqnarray}

\noindent We differentiate this equation with respect to $\rho$ and
combine it with Eq. (2.7) to obtain

\begin{eqnarray}
\frac{\partial}{\partial\rho} G^{(n)}(p_1,...,p_n;\overline{g}(\rho),
\overline{\mu}(\rho)) &=&
R(\rho)\,{\cal D}G^{(n)}(p_1,...,p_n;g,\mu)\nonumber\\ &=&
-nR(\rho)\gamma_{\phi}(\rho)~G^{(n)}(p_1,...,p_n;g,\mu)\nonumber\\ &=&
-n\gamma_{\phi}(\overline{g}(\rho))~G^{(n)}(p_1,...,p_n;\overline{g}(\rho),
\overline{\mu}(\rho))\per .
\label{2.17}\end{eqnarray}

\noindent We have to introduce a boundary condition to this
differential equation. In a cut-off theory we may set 

\begin{eqnarray}
\lim_{\mu\rightarrow \infty}G^{(n)}(p_1,...,p_n;g(\mu),\mu)=
G^{(n)}_0(p_1,...,p_n;g_0)\per ,
\label{2.18}\end{eqnarray}

\noindent where $g_0$ denotes the bare coupling constant.

By integrating Eq. (2.17) we find

\begin{eqnarray}
G^{(n)}(p_1,...,p_n;g,\mu)= exp\left[n\int_0^{\rho}d\rho
\gamma_{\phi}(\overline{g}(\rho))\right]\cdot 
G^{(n)}(p_1,...,p_n;\overline{g}(\rho),
\overline{\mu}(\rho))\per .
\label{2.19}\end{eqnarray}

\noindent In the limit $\rho\rightarrow \infty$ and consequently
$\overline{\mu}(\rho)\rightarrow \infty$ we have

\begin{eqnarray}
G^{(n)}(p_1,...,p_n;g,\mu)=  exp\left[n\int_0^{\infty}d\rho
\gamma_{\phi}(\overline{g}(\rho))\right] \cdot G^{(n)}_{0}(p_1,...,p_n;g_0)
\per .
\label{2.20}\end{eqnarray}

\noindent In a cut-off theory all the vertex corrections to $g(\mu)$
for $\mu\rightarrow \infty$ tend to vanish leaving only the bare one,
namely,

\begin{eqnarray}
\lim_{\mu\rightarrow \infty}g(\mu)= \lim_{\rho\rightarrow \infty}
\overline{g}(\rho)= g_0\per .
\label{2.21}\end{eqnarray}

\noindent The fundamental field $\phi$ is multiplicatively
renormalized as

\begin{eqnarray}
\phi^{(0)}(x) = Z^{1/2}_{\phi}\phi (x)\per ,
\label{2.22}\end{eqnarray}

\noindent where $Z_{\phi}$ is the renormalization constant of the
scalar field $\phi$, and it is a function of $g$. Comparison of
Eqs. (2.20) and (2.22) yields

\begin{eqnarray}
Z^{-1}_{\phi}= exp\left[2\int_0^{\infty} d\rho
\gamma_{\phi}(\overline{g}(\rho))\right]\per .
\label{2.23}\end{eqnarray}

\noindent The running renormalization constant is given by

\begin{eqnarray}\begin{array}{l}
Z^{-1}_{\phi}(\rho)= R(\rho)Z^{-1}_{\phi}(g)\\
\\
=exp\left[2\int_{\rho}^{\infty} d\rho^{\prime}
\gamma_{\phi}(\overline{g}(\rho^{\prime}))\right]\per .\\\end{array}
\label{2.24}\end{eqnarray}

\noindent When $Z_{\phi}$ depends not only on $g$ but also on $\mu$,
$\gamma_{\phi}(\rho)$ must be replaced by $\gamma_{\phi}(\rho,\mu)$.

In {\it a cut-off theory} we have 

\begin{eqnarray}
\lim_{\rho\rightarrow \infty}Z^{-1}_{\phi}(\rho)=1\per ,
\label{2.25}\end{eqnarray}

\noindent but this is not true when the integral in the exponent of
Eq.  (2.23) does not converge and as we shall see later this
feature is a possible cause of the emergence of the GIS terms.

Although the RG approach has been introduced for the scalar theory we
can easily extend it to gauge theories.  In QED the generator of the
RG is given by

\begin{eqnarray}
{\cal D}= \mu\frac{\partial}{\partial \mu} +\beta (e)
\frac{\partial}{\partial e} -2\alpha \gamma_V(e)
\frac{\partial}{\partial \alpha}\per ,
\label{2.26}\end{eqnarray}

\noindent where $\alpha$ denotes the gauge parameter.  The $\gamma_V(e)$
denotes the  anomalous dimension of the electromagnetic field and is
related to $\beta(e)$ through the Ward identity 

\begin{eqnarray}
\beta (e) = e\gamma_V (e)\per .
\label{2.27}\end{eqnarray}

\noindent Furthermore in QCD the generator is given by

\begin{eqnarray}
{\cal D}= \mu\frac{\partial}{\partial\mu} +\beta (g)
\frac{\partial}{\partial g} -2\alpha \gamma_V(g,\alpha)
\frac{\partial}{\partial \alpha}\per ,
\label{2.28}\end{eqnarray}

\noindent where $g$ denotes the gauge coupling constant and $\gamma_V$
the anomalous dimension of the color gauge field. The running
parameters in QCD satisfy the following equations:

\begin{eqnarray}
\frac{d\overline{g}}{d\rho}&=& \beta (\overline{g})\per ,
\label{2.29}\\
\frac{d\overline{\alpha}}{d\rho}&=& -2 \overline{\alpha} 
\gamma_V(\overline{g}, \overline{\alpha})
\per .
\label{2.30}
\end{eqnarray}

\noindent Then we introduce their asymptotic values by 

\begin{eqnarray}
g_{\infty}= \lim_{\rho\rightarrow \infty}\overline{g}(\rho)\per , 
\per \per \alpha_{\infty}= \lim_{\rho\rightarrow
\infty}\overline{\alpha}(\rho)\per .\label{2.31}
\end{eqnarray}

\noindent This is possible since the RG is a group $GL(1,R)$ but not
$U(1)$.  Asymptotic freedom$^{13,14)}$ of QCD implies

\begin{eqnarray}
g_{\infty}=0\per .
\label{2.32}\end{eqnarray}

\noindent By integrating Eq. (2.30) we immediately find a sum
rule, 

\begin{eqnarray}
2\int_0^{\infty} d\rho \overline{\gamma}_{V}(\rho)=
ln(\frac{\alpha}{\alpha_{\infty}})
\label{2.33}\end{eqnarray}

\noindent and hence  we also have$^{4,5,12)}$

\begin{eqnarray}
Z_3^{-1} = exp\left[2\int_0^{\infty} d\rho \overline{\gamma}_{V}(\rho)\right]=
\frac{\alpha}{\alpha_{\infty}}\per ,
\label{2.34}\end{eqnarray}

\noindent where 
$\overline{\gamma}_V(\rho)\equiv \gamma_V(\overline{g}(\rho),
\overline{\alpha}(\rho))$.

In QCD it is known that $\alpha_{\infty}$ can take three possible
values$^{4,5,12)}$

\begin{eqnarray}
\alpha_{\infty}= 0\per ,\per  \alpha_0\per ,\per  -\infty\per ,
\label{2.35}\end{eqnarray}

\noindent where $\alpha_0$ is a constant which depends only on the
number of quark flavors. These three values are related to the
integral of $\gamma_V$ as 

\begin{eqnarray}
\int_0^{\infty} d\rho \overline{\gamma}_{V}(\rho)= 
\left\{   \begin{array}{ll}
\infty\per , & \mbox{for}\per  \alpha_{\infty}=0\\ 
\mbox{finite}\per ,& \mbox{for} \per \alpha_{\infty}=\alpha_0\\ 
-\infty\per ,&\mbox{for}\per  \alpha_{\infty}=-\infty\\
\end{array}\right.
\label{2.36}\end{eqnarray}

\noindent and $Z_3^{-1}$ vanishes when $\alpha_{\infty}=-\infty$.

\section{Quantum Electrodynamics}

Quantum electrodynamics is a suitable ground to exercise the analysis
of the  GIS terms. The Lagrangian density for QED is given by

\begin{eqnarray}
{\cal L} = {\cal L}_{em}+ {\cal L}_{matter}\per ,
\label{3.1}\end{eqnarray}

\noindent where the unrenormalized version of the Lagrangian density
for the electromagnetic field is given by

\begin{eqnarray}
{\cal L}_{em}= -\frac{1}{4}F_{\mu\nu}^{(0)}\cdot 
F_{\mu\nu}^{(0)} + \partial_{\mu}
B^{(0)}\cdot A_{\mu}^{(0)} +\frac{\alpha_0}{2} B^{(0)}\cdot B^{(0)}\per ,
\label{3.2}\end{eqnarray}

\noindent where $B$ denotes the Nakanishi-Lautrup auxiliary field$^{10)}$ 
and
the interactions are included in the matter Lagrangian. The resulting
field equations are given by 

\begin{eqnarray}
\partial_{\mu}F_{\mu\nu}^{(0)} +\partial_{\nu} B^{(0)}&=& -J_{\nu}^{(0)}\per ,
\label{3.3}\\
\nonumber\\
  \partial_{\mu}A_{\mu}^{(0)}&=& \alpha_0 B^{(0)}\per ,
\label{3.4}\end{eqnarray}

\noindent and the renormalized version of these equations can be
expressed as 

\begin{eqnarray}\begin{array}{rl}
\partial_{\mu}F_{\mu\nu} +\partial_{\nu} B=& -J_{\nu}\per ,
\end{array}
\label{3.5}\\ 
\nonumber\\
\begin{array}{rl}
 \partial_{\mu}A_{\mu}=& \alpha B\per .
\end{array}
\label{3.6}\end{eqnarray}

\noindent The fundamental fields $A_{\mu}$ and $B$ as well as the
gauge parameter $\alpha$ are renormalized multiplicatively,

$$\begin{array}{llr}
\hspace{5.0cm}&A_{\mu}^{(0)}= Z_3^{1/2}A_{\mu} \per , & \hspace{5.2cm} (3.7a) \\
&\\
\hspace{5.0cm}&B^{(0)}= Z_3^{-1/2}B \per ,   & \hspace{5.2cm} (3.7b)\\ 
&&\\
\hspace{5.0cm}&\alpha_0=Z_3\alpha\per .   &  
\hspace{5.2cm}(3.7c)\\
\end{array}$$
\addtocounter{equation}{+1}

\noindent Apparently renormalization of the composite current operator
 $J_{\nu}$
is not multiplicative, but its execution requires operator
mixing$^{3,5)}$ as illustrated by

$$\begin{array}{lrr}
\hspace{4.0cm}&J_{\nu}^{(0)}= Z_3^{1/2}\left[
J_{\nu}+(1-Z_3^{-1})\partial_{\nu}B\right] \per ,  &\hspace{3.2cm}(3.8a)\\
\mbox{or}& & \\
\hspace{4.0cm}&J_{\nu}= Z_3^{-1/2}\left[ J_{\nu}^{(0)}+(1-Z_3)\partial_{\nu}B^{(0)}\right]
\per .& \hspace{3.2cm}(3.8b)\\
\end{array}$$

\addtocounter{equation}{+1}

\noindent Operator mixing is one of the sources of the GIS terms, and
in order to illustrate this statement we shall evaluate the ETC

\begin{eqnarray}
\delta (x_0-y_0)\left[ A_j(x), \per J_0(y)\right]
\label{3.9}\end{eqnarray}

\noindent for $j=1,2,3$. In the unrenormalized version we have 

\begin{eqnarray}
\delta (x_0-y_0)\left[ A_j^{(0)}(x),\per  J_0^{(0)}(y)\right]=0\per .
\label{3.10}\end{eqnarray}

\noindent As has been mentioned before we can rely on the ETCs only
between two fundamental fields, so that we shall express $J$ in terms
of $A$ and $B$ by using Eqs. (3.5) and (3.7),

$$\begin{array}{rl}
\left[ A_j(x), \per J_4(y)\right]=
 &-\left[A_j(x), \per \partial_{\mu}
F_{\mu4}(y)+ \partial_4B(y)\right]\\ 
\\
=& -Z_3^{-1}
\left[A_j^{(0)}(x),\per  \partial_{k} F_{k4}^{(0)}(y)\right]-  
\left[A_j^{(0)}(x),\per\partial_4B^{(0)}(y) \right]\\
\\
 =& (-Z_ 3^{-1} +1) \partial_j
\delta^3(x-y)\\
\end{array}$$

\noindent for $x_0=y_0$. Thus we have 

\begin{eqnarray}
\begin{array}{rl}
\delta (x_0-y_0)\left[ A_j(x),\per J_0(y)\right]=& i (Z_ 3^{-1} -1)
\partial_j \delta^4(x-y)\\ 
\\
\equiv&is \partial_j \delta^4(x-y)\per ,\\
\end{array}
\label{3.11}\end{eqnarray}

\noindent where $s$ is the coefficient of the GIS term. In this case
it is clear that the origin of the GIS term is the operator
mixing. Then $s$ satisfies the RG equation

\begin{eqnarray}
\left[{\cal D}+ 2\gamma_V(e)\right](s+1)=\left[{\cal D}+
2\gamma_V(e)\right]Z_3^{-1} =0\per , \label{3.12}\end{eqnarray}

\noindent where ${\cal D}$ is given by Eq. (2.26), and the running GIS
coefficient $\overline{s}(\rho)$ satisfies the differential equation

\begin{eqnarray}
\left[\frac{\partial}{\partial\rho}+
2\gamma_V(\overline{e}(\rho))\right] (\overline{s}(\rho)+1)=0\per .
\label{3.13}\end{eqnarray}

\noindent In a cut-off theory the GIS term is absent in the
unrenormalized version as expressed by Eq. (3.10), and the
boundary condition for  $\overline{s}(\rho)$ is given by

\begin{eqnarray}
\overline{s}(\infty)= 0\per .
\label{3.14}\end{eqnarray}

\noindent By combining the boundary condition (3.14) with
Eq. (3.13) we find the solution 

\begin{eqnarray}
Z_3^{-1}(\rho)= 1+ \overline{s}(\rho)= exp\left[2\int_{\rho}^{\infty}
d\rho^{\prime} \gamma_V(\overline{e}(\rho^{\prime})) \right]\per .
\label{3.15}\end{eqnarray}

\noindent In  the absence of the cut-off we do not know what kind of
boundary condition we should impose on $\overline{s}(\rho)$  so that we
take this solution (3.15) for granted even in this case.

In QED we assume that $Z_3^{-1}=Z_3^{-1}(0)$ is divergent so that we
have 

\begin{eqnarray}
1+ \overline{s}(\infty)= \lim_{\rho\rightarrow
\infty}exp\left[2\int_{\rho}^{\infty} d\rho^{\prime}
\gamma_V(\overline{e}(\rho^{\prime})) \right]=\infty\per , 
\label{3.16}\end{eqnarray}

\noindent and the boundary condition (3.14) is no longer satisfied
in the absence of the cut-off. This is another source of the GIS
terms, and the field operators do not necessarily tend to the
unrenormalized ones in the limit $\rho\rightarrow \infty$ and hence
$\mu\rightarrow \infty$ when the cut-off is lifted.

 Finally we shall turn our attention to the ETC between two components
 of the current density. This is precisely the original problem in
 which the GIS term was recognized$^{1,2)}$. We shall make use of the
 field equations (3.5) to express the current density as a linear
 combination of the fundamental fields, and then we can make use of
 the commutativity of $B$ with $F_{\mu\nu}$ and $B$ itself$^{10)}$,

\begin{eqnarray}\begin{array}{rl}
\left[J_{\mu}(x),\per J_{\nu}(y) \right]=&
\left[\partial_{\alpha}F_{\alpha\mu}(x)+\partial_{\mu} B(x),\per
\partial_{\beta}F_{\beta\nu}(y)+\partial_{\nu} B(y) \right]\\
\\
=&\left[\partial_{\alpha}F_{\alpha\mu}(x),\per
\partial_{\beta}F_{\beta\nu}(y) \right]\per ,\\
\end{array}
\label{3.17}\end{eqnarray}

\noindent and we introduce the GIS coefficient $s$ by 

\begin{eqnarray}
\delta (x_0-y_0)\langle 0\vert  \left[J_{j}(x),\per J_{0}(y) \right] \vert 0
  \rangle= is \partial_j\delta^4 (x-y)\per .
\label{3.18}\end{eqnarray}

\noindent  As a matter of fact, the ETC on the left-hand-side of
Eq. (3.18) is known to be a c-number before taking its
vacuum expectation value in spinor electrodynamics$^{15)}$.
Here we are aware of the fact that the GIS term can be
expressed in terms of the ETC between derivatives of field
strengths. In order to evaluate the ETC by making use of the CCRs it
is necessary to express derivatives of field strengths in terms of
canonical variables by making use of canonical field
equations. Therefore, we are taking commutators between those operators
that are non-local in time and then taking the local limit. The way in
which this limit is taken is dictated in the evaluation of the higher
order corrections as we shall see below. 

This is in a sharp contrast to the original naive way of evaluating
the commutator between two bilinear forms of the Dirac fields by
making use of only the CCRs without taking the possibility
of non-locality into consideration. This
gap generates the GIS term.

By combining Eqs. (3.17) and (3.18) we find that the GIS
coefficient $s$ satisfies the RG equation 

\begin{eqnarray}
\left[{\cal D}+ 2\gamma_V(e)\right]s=0\per .
\label{3.19}\end{eqnarray}

\noindent In this case we cannot give the boundary condition for this
equation since it requires the information about the dynamics of the
system such as the photon propagator.  The Lehmann representation of
the electromagnetic field is given in the following form:

\begin{eqnarray}
\langle 0\vert T\left[A_{\mu}(x), A_{\nu}(y)  \right]\vert 0\rangle =
\frac{-i}{(2\pi)^4}\int d^4k e^{ik\cdot (x-y)} D_{F\mu\nu}(k)\per
, \label{3.20}\\
\nonumber\\
D_{F\mu\nu}(k)=
\left(\delta_{\mu\nu}-\frac{k_{\mu}k_{\nu}}{k^2-i\epsilon}
\right)\int dM^2 \frac{{\Large \rho}(M^2;e,\mu)}{k^2+ M^2-i\epsilon} +
 \alpha \frac{k_{\mu}k_{\nu}}{(k^2-i\epsilon)^2}\per . \label{3.21}
\end{eqnarray}

\noindent Then inserting this expression into Eq. (3.17) we find 

\begin{eqnarray}
s= \int dM^2 {\Large \rho}(M^2;e,\mu)M^2\per .
\label{3.22}\end{eqnarray}

\noindent This expression certainly satisfies Eq. (3.19) since we
have

\begin{eqnarray}
\left[{\cal D}+ 2\gamma_V(e)\right]{\Large \rho}(M^2;e,\mu)=0\per .
\label{3.23}\end{eqnarray}

\noindent It is clear that $Z_3^{-1}$ also satisfies Eq. (3.19) since it
is given by 

\begin{eqnarray}
Z_3^{-1}= \int dM^2 {\Large \rho}(M^2;e,\mu)\per .
\label{3.24}\end{eqnarray}

\noindent We may conclude that the GIS terms are controlled by RG if
not completely.

\section{Color Confinement in Quantum Chromodynamics}

 In QCD the GIS term plays an important role in connection with color
confinement$^{3-5)}$. The field equation in QCD corresponding to
Eq. (3.5) is given by 

\begin{eqnarray}
\partial_{\mu}F_{\mu\nu}^{a} + J_{\nu}^a= i\delta\overline{\delta}
 A_{\nu}^a \per ,
\label{4.1}\end{eqnarray}

\noindent where $\delta$ and $\overline{\delta}$  denote two kinds of
Becchi-Rouet-Stora (BRS) transformations$^{11)}$, respectively, and the
superscript $a$ represents the color index. Since we are not entering
the subject of BRS transformations here we shall refer to other
references$^{3-5)}$ for their definitions.

We are interested in ETC 

$$\begin{array}{l} \partial_{\mu}\langle 0\vert
T\left[i\delta\overline{\delta}A_{\mu}^a(x), \per A_{j}^b(y)
\right]\vert 0\rangle\\
\\
 =\delta (x_0-y_0) \langle 0\vert
\left[i\delta\overline{\delta}A_{0}^a(x),\per A_{j}^b(y)
\right]\vert 0\rangle\\ 
\\
= i\delta_{ab}C~ \partial_j\delta^4 (x-y)\per
,\\\end{array}$$

\noindent or

\begin{eqnarray}
\delta (x_0-y_0) \langle 0\vert \left[\partial_kF_{k4}^a(x) +J_4^a(x),
 \per A_{j}^b(y) \right]\vert 0\rangle \nonumber\\
\nonumber\\
 = -\delta_{ab}C~
\partial_j\delta^4 (x-y)\per .
\label{4.2}\end{eqnarray}

\noindent The constant $C$ is gauge-dependent, and a sufficient
condition for color confinement is the existence of a gauge in which
the following equality holds:

\begin{eqnarray}
C=0\per .
\label{4.3}\end{eqnarray}

\noindent In order to determine $C$ we have to evaluate the ETC in Eq.
 (4.2), and for that purpose we introduce the RG equation satisfied
 by  $C~^{3-5)}$,

\begin{eqnarray}
\left({\cal D}-2\gamma_{FP} \right)C=0\per ,
\label{4.4}\end{eqnarray}

\noindent where ${\cal D}$ is given by Eq. (2.28) and $\gamma_{FP}$ denotes
the anomalous dimension of the Faddeev-Popov ghost fields. Then the
renormalization constant of the ghost fields denoted by $\tilde{Z}_3$ also
satisfies the same RG equation,

\begin{eqnarray}
\left({\cal D}-2\gamma_{FP} \right){\tilde Z}_3=0\per .
\label{4.5}\end{eqnarray}

\noindent We are going to study the relationship between $C$ and $\tilde{Z}_3$
in this section. They satisfy the same RG equation, but their
normalizations are different.

The unrenormalized version of Eq. (4.1) reads as

\begin{eqnarray}
i\delta\overline{\delta}A_{\nu}^{(0)}(x)= \partial_{\mu}A^{(0)}_{\mu\nu} +
g_0 \partial_{\mu}(A_{\mu}^{(0)}\times A_{\nu}^{(0)})
+J_{\nu}^{(0)}\per ,
\label{4.6}\end{eqnarray}

\noindent where $A_{\mu\nu}=
\partial_{\mu}A_{\nu}-\partial_{\nu}A_{\mu}$ denotes the linear part
of $F_{\mu\nu}$ and we have suppressed the color index. The cross
product denotes the antisymmetric product in the color space defined
in terms of the structure constants of the algebra $su(3)$. When we
insert the r.h.s. of Eq. (4.6) into the ETC (4.2) in the
unrenormalized version, we find that only the first term
$\partial_{\mu}A_{\mu\nu}^{(0)}$ gives a non-vanishing canonical commutator
and the rest would give only a  vanishing result provided that the
naive CCRs are employed. However, this is true only in a cut-off
theory or in a convergent theory and in general we should not discard
the possibility of a non-vanishing GIS term so that the unrenormalized
constant $C_0$ would be given by 

\begin{eqnarray}
C_0=1+s\per .
\label{4.6a}\end{eqnarray}

\noindent The first term is a result of the CCR and is equal to
unity. Thus the renormalized $C$ is given by

\begin{eqnarray}
C= C_0{\tilde  Z}_3 =(1+s){\tilde Z}_3\per .
\label{4.7}\end{eqnarray}

\noindent Then a   question is raised of how to evaluate the GIS
coefficient $s$. For this purpose we introduce a cut-off theory and
we write  $\overline{a}(\rho)$ for $Z_3^{-1}(\rho)$, and we shall rewrite
Eq. (4.7) in the form

\begin{eqnarray}
\overline{C}(\infty)= \overline{a}(\infty)
\label{4.8}\end{eqnarray}

\noindent based on the argument developed in Sec. 2. In a cut-off
theory the GIS coefficient $s$ vanishes, but it does not vanish when
the cut-off is lifted.  The running parameters $\overline{C}(\rho)$,
$\overline{a}(\rho)$ and ${\tilde Z}_3(\rho)$ satisfy the following
differential equations, respectively, 

\begin{eqnarray}
\left[ \frac{\partial}{\partial{\rho}}-2\overline{\gamma}_{FP}(\rho)
\right]\overline{C}(\rho)=0\per ,\label{4.9}\\ 
\nonumber\\
 \left[
\frac{\partial}{\partial{\rho}}+2\overline{\gamma}_{V}(\rho)
\right]\overline{a}(\rho)=0\per ,\label{4.10}\\
\nonumber\\
 \left[
\frac{\partial}{\partial{\rho}}-2\overline{\gamma}_{FP}(\rho)
\right]{\tilde Z_3}(\rho)=0\per .\label{4.11}  
\end{eqnarray}

\noindent Among them the last two are renormalization constants, and
they are immediately given by

\begin{eqnarray}
\overline{a}(\rho)= Z_3^{-1}(\rho)= exp\left[2\int_{\rho}^{\infty}d\rho^{\prime}
\overline{\gamma}_{V}(\rho^{\prime})\right]\per ,\label{4.12}\\
\nonumber\\
{\tilde  Z}_3^{-1}(\rho)=
exp\left[2\int_{\rho}^{\infty}d\rho^{\prime} \overline{\gamma}_{FP}(\rho^{\prime})\right]\per
.\label{4.13}
\end{eqnarray}

\noindent We should be aware of the following relations:

\begin{eqnarray}
\begin{array}{ll}
Z_3^{-1}=Z_3^{-1}(0)\per , & {\tilde Z}_3^{-1}={\tilde Z}_3^{-1}(0)\per ,\\
\end{array} \label{4.14}
\end{eqnarray}

\noindent Then $\overline{C}(\rho)$ should be determined by solving Eq.
(4.10) under the boundary condition (4.9) and we obtain

\begin{eqnarray}
\overline{C}(\rho)= \lim_{\rho^{\prime}\rightarrow \infty} exp\left[ 2
\int_{\rho^{\prime}}^{\infty}d\rho^{\prime\prime}
\overline{\gamma}_{V}(\rho^{\prime\prime}) - 2
\int_{\rho}^{\rho^{\prime}}d\rho^{\prime\prime}
\overline{\gamma}_{FP}(\rho^{\prime\prime})  \right]\per ,
\label{4.15}\end{eqnarray}

\noindent and, in particular, we have

\begin{eqnarray}
C= \lim_{\rho^{\prime}\rightarrow \infty} exp\left[ 2
\int_{\rho^{\prime}}^{\infty}d\rho^{\prime\prime}
\overline{\gamma}_{V}(\rho^{\prime\prime}) - 2
\int_{0}^{\rho^{\prime}}d\rho^{\prime\prime}
\overline{\gamma}_{FP}(\rho^{\prime\prime})  \right]\per .
\label{4.16}\end{eqnarray}

\noindent From now on we lift the cut-off while keeping these formulas.
With recourse to Eqs. (2.34) and (2.36) we find that $C$
vanishes when $Z_3^{-1}$ vanishes as claimed before$^{3-5)}$.  Then we
may express Eq. (4.17) as 

\begin{eqnarray}
C= \lim_{\rho \rightarrow \infty} exp\left[ 2 \int_{\rho}^{\infty}d\rho^{\prime}
\overline{\gamma}_{V}(\rho^{\prime}) \right]\cdot {\tilde Z}_3\per ,
\label{4.17}\end{eqnarray}

\noindent and with reference to Eq. (4.8) we find

\begin{eqnarray}
1+s = \lim_{\rho \rightarrow \infty} exp\left[ 2
\int_{\rho}^{\infty}d\rho^{\prime} \overline{\gamma}_{V}(\rho^{\prime})
\right]\nonumber\\
\nonumber\\
 = \left\{ \begin{array}{ll} \infty\per ,& \mbox{for} \per
\alpha_{\infty}=0\\ 1\per , & \mbox{for} \per \alpha_{\infty}=\alpha_0\\
0\per , & \mbox{for} \per \alpha_{\infty}=-\infty
\end{array}\right.
\label{4.18}\end{eqnarray}

\noindent Only in the case $\alpha_{\infty}=\alpha_{0}$  do we find
the vanishing GIS coefficient $s$, and this is precisely what happens
when the integration of $\overline{\gamma}_V$ converges just as in the
cut-off theory. Now we shall summarize the relationship between $C$
and ${\tilde Z}_3$ as follows:

\begin{eqnarray}
C= \left\{\begin{array}{ll} \infty\per , &\alpha_{\infty}=0\\ 
{\tilde Z}_3\per , &
\alpha_{\infty}=\alpha_0\\ 0 & \alpha_{\infty}=-\infty\\
\end{array}\right.\label{4.19} \end{eqnarray}

As we have seen above we formulate the boundary condition
for a given RG
equation by introducing a cut-off, but when the cut-off is lifted in
the solution the GIS term appears as a manifestation of the divergent
character of the theory.

Acknowledgement: The financial support of the Academy of Finland under 
the Project no. 163394 is greatly acknowledged.

\end{document}